\documentclass{optica-article}

\journal{opticajournal} 

\articletype{Research Article}

\usepackage{lineno}

\begin{document}

\title{Wavelength-diverse transmission for turbulence-resistant free-space optical communication}

\author{Nicolas~Couture\authormark{*}, Brandon~Buscaino, Douglas~Charlton, Mohammad~E.~Mousa-Pasandi, and Kim~B.~Roberts}

\address{Ciena Corporation, Ottawa, ON K2K 0L1, Canada}
\email{\authormark{*}ncouture@ciena.com} 

\begin{abstract*} 
Free-space optical communication~(FSOC) links are susceptible to outages caused by atmospheric turbulence-induced fading. Wavelength diversity can mitigate fading by transmitting correlated information across substantially uncorrelated channels. In this study, we demonstrate a wavelength diversity system to mitigate the effects of turbulence on a signal propagated in free space. A modified polarization-multiplexed coherent optical transceiver transmits a 56.8~GBaud signal at a client rate of 200~Gbps across two carrier frequencies within the C-band over a bench-top free-space link with a turbulence emulator. Two synchronized single-wavelength coherent receivers capture the received waveforms at each carrier frequency, which are subsequently digitally combined using maximal ratio combining for offline processing. Our results demonstrate a reduction in outage probability by nearly a factor of 20 compared to the single carrier transmission scenario. Finally, we observe fading-dependent correlation between wavelengths in the C-band, which is exploited to reduce the outage probability by 86\%.
\end{abstract*}

\section{Introduction}
In recent years, optical frequencies have been increasingly utilized in free-space communication links, replacing traditional radio frequencies (RF) to address the rapidly growing demand for client capacity. Optical frequencies offer higher bandwidths and data rates compared to RF, making them ideal for modern communication systems. Remote areas underserved by traditional telecommunications infrastructure may be incorporated into a global communications network with low-earth orbit (LEO) satellites. This connectivity between geographically isolated regions and the broader network infrastructure could be enabled by high-speed free-space optical communications (FSOC)~\cite{schieler2023orbit}.\\
FSOC links face a range of complex technical challenges and environmental constraints that limit their widespread adoption. Advanced control algorithms are required to ensure that the transmitted optical beam profile sufficiently and coherently aligns with the receiver aperture. From an environmental perspective, severe weather conditions such as heavy rain, snow, or fog can temporarily disrupt FSOC links, necessitating the use of slower backup RF links, which leverage lower carrier frequencies. Another critical challenge is clear-air turbulence, caused by air movement, temperature gradients, and pressure fluctuations, which can significantly distort optical signals. This turbulence often results in power fading, where multi-path interference reduces the received power below the receiver's sensitivity threshold, further complicating reliable FSOC performance.\\
Turbulence mitigation techniques have been extensively studied, each with distinct advantages and limitations. One solution involves the use of adaptive optics at the transmitter and/or receiver ends~\cite{Tyson2022}. In these systems, the optical field is measured by a wavefront sensor and subsequently pre- or post-compensated using a deformable mirror or spatial light modulator, enabling effective transmission over turbulent classical~\cite{wang_performance_2018} and even quantum~\cite{Scarfe_Hufnagel_Ferrer-Garcia_DErrico_Heshami_Karimi_2025} FSOC links. While highly effective, adaptive optics systems require complex hardware, precise calibration, and substantial computational resources, making their implementation challenging.\\
Other turbulence mitigation techniques leverage diversity, transmitting redundant information across multiple partially- or fully-uncorrelated channels to minimize bit errors. These methods hold significant promise for FSOC systems due to their high signal gain and resistance to atmospheric fading. Temporal diversity, for example, involves transmitting identical information on time-delayed subcarriers~\cite{Popoola2012}, effectively mitigating errors caused by turbulence-induced fading. Similarly, spatial diversity schemes utilize multiple transmitter and/or receiver apertures to exploit the spatial decorrelation of atmospheric turbulence, thereby reducing the probability of link outages~\cite{Navidpour2007}.\\
In this work, we utilize wavelength diversity by encoding identical information on different wavelengths that share the same optical path~\cite{Abadi2015,Niu2013,Purvinskis2003,rachmani2010wavelength,shah2021intl,Nafria2024}. While previously reported wavelength diversity schemes typically operate over bandwidths spanning several hundred nanometers to achieve strong decorrelation, we utilize wavelengths at the edges of the C-band. This approach provides moderate decorrelation~\cite{weerackody2006wavelength} while ensuring compatibility with readily available optical components and existing infrastructure, making it a practical and efficient solution. Here, a coherent transceiver is modified to simultaneously transmit two wavelengths, delivering 200~Gbps at a symbol rate of 56.8~GBaud across a bench-top free-space link with a turbulence emulator. Our experimental results demonstrate a reduction of the outage probability by a factor of approximately 20 compared to single wavelength transmission. We also measure fading-dependent correlation between wavelengths, which can be exploited to further improve the resilience of wavelength-diverse systems. These improved wavelength-diverse solutions are more tolerant to deep power fades, offering enhanced reliability and resilience in challenging atmospheric conditions.\\
\section{Theory}
\begin{figure}[b]
\centering\includegraphics[width=\columnwidth]{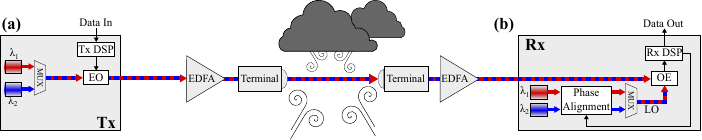}
\caption{Schematic of wavelength diverse free-space transmission where both wavelengths a) share their electro-optics~(EO) at the transmitter~(Tx) and b) share their opto-electronics~(OE) at the receiver~(Rx). An additional per-wavelength phase alignment step is used on the LO lasers to facilitate maximal ratio combining~(MRC) within the shared OE.}
\label{fig:FSOCLink}
\end{figure}
\begin{figure}[h]
    \centering
    \includegraphics[width=0.55\columnwidth]{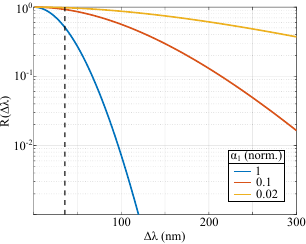}
    \caption{Two-frequency mutual coherence function described by Eq.~\ref{Eq:CoherenceFunction} for various levels of turbulence. The black dashed line represents the 35~nm channel separation relevant to this work.}
    \label{fig:R}
\end{figure}

An ideal wavelength-diverse FSOC link architecture is shown in Fig.~\ref{fig:FSOCLink}. The depicted configuration utilizes a single transmitter~(Tx) and a single receiver~(Rx), where two wavelength-detuned lasers are multiplexed within the modem. These two wavelengths undergo the same modulation in the Tx electro-optics~(Fig.~\ref{fig:FSOCLink}a), ensuring that identical information is encoded on each wavelength. In the Rx~(Fig.~\ref{fig:FSOCLink}b), these two laser tones serve as a pair of local oscillators for demodulation in the opto-electronics. To harness the benefit of wavelength-diversity, an additional feedback loop would be implemented in the Rx hardware to align the phase of the lasers to the received signal~\cite{US18541189}. This step ensures that the currents add coherently after the optical-to-electrical signal conversion with the proper per-wavelength weighting, allowing for maximal ratio combining~(MRC) of the two waves.\\ 
In Kolmogorov’s theory of turbulence, it is assumed that the atmosphere is composed of eddies of varying sizes~\cite{kolmogorov1941local}. The distribution of eddy sizes which, along with other parameters such as wind speed and pressure, determine the cumulative effects of refraction and interference of an optical signal as it traverses the atmosphere. Additional models have been developed using the foundation of Kolmogorov’s theory, such as the Gamma-Gamma model whose probability density function is given by
\begin{equation}
    p(I) = \frac{2(\alpha\beta)^{(\alpha+\beta)/2}}{\Gamma(\alpha)\Gamma(\beta)}I^{(\alpha+\beta)/2 -1}K_{\alpha-\beta}[2\sqrt{\alpha\beta I}]
    \label{eq:gammagamma}
\end{equation}
where $I$ is the irradiance, $\Gamma$ is the Gamma function, and $K$ is the second-order Bessel function~\cite{alhabash2001irradiance}. Atmospheric turbulence in the Gamma-Gamma model is characterized primarily by the size of the smallest ($\alpha$) and largest ($\beta$) atmospheric eddies. These length scales are then used to predict the distribution of the intensities at the receiver plane. In the receiver plane, the coherence length, $r_0 = 1.68C_n^2Lk^2$, quantifies the spatial coherence for an incident plane wave after propagation through a turbulent atmosphere~\cite{fried1967optical,andrews2005laser}. Here, $L$ is the link length, $k=2\pi/\lambda$ is the propagation constant and $C_n^2$ represents the turbulence strength. The two-frequency mutual coherence function for coherently downconverted optical signals, such as those recovered in a typical coherent optical receiver, is plotted in Fig.~\ref{fig:R} for various levels of turbulence, is
\begin{equation}
    R(\lambda_1,\lambda_2)\approx \exp(-2\pi^2\alpha_1(k_1-k_2)^2)
    \label{Eq:CoherenceFunction}
\end{equation}
where $\alpha_1$ is a constant proportional to the turbulence strength, $C_n^2$, and $k_{i=1,2}$ are the propagation constants of the two frequencies of interest~\cite{kelly1999temporal,weerackody2006wavelength}. For two frequencies within the conventional optical band~(C-band), as demonstrated in this work, a maximum wavelength difference of $\sim$35~nm is achievable, resulting in $R$ greater than 0.9 for most turbulence strengths.\\
\section{Methods}
\subsection{Experimental Setup}
\begin{figure}[h]
\centering\includegraphics[width=\columnwidth]{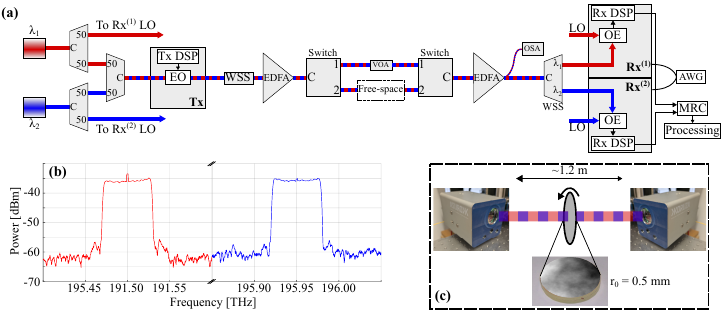}
\caption{(a) Experimental configuration. (b) Output spectrum of the modified coherent transceiver, where two 56.8~GHz signals are simultaneously produced at carrier frequencies of 191.5 and 195.95~THz sharing the same electro-optics. (c) Schematic of the free-space channel comprised of two fiber-to-free-space terminals transmitting and receiving the dual-wavelength coherent signal, and a rotating phase plate emulating atmospheric turbulence.}
\label{fig:Setup}
\end{figure}

The experimental setup is shown in Fig.~\ref{fig:Setup}a. Two micro-integrable tunable laser assembly~(uITLA) sources are centered at 191.5~THz~(red) and 195.95~THz~(blue), respectively, each with an output power of +17~dBm. Each laser is split with a 50/50 coupler. One path is used as the local oscillator~(LO) in receiver Rx$^{(1)}$ or Rx$^{(2)}$. The other paths are combined to generate a dual-wavelength source. The transmitter consists of a coherent modem, with sampling frequency of 68.16~GHz, modified to accept an external laser source. The dual-wavelength laser source is fed into the transmitter to generate two 56.8~GHz dual-polarization channels (Fig.~\ref{fig:Setup}b), each with bit rates of 200~Gbps. A wavelength-selective switch (WSS) is used to balance the power between channels on the transmitter end. An optical switch allows us to alternate between all-fiber (path~1 in Fig.~\ref{fig:Setup}a) and free-space (path~2 in Fig.~\ref{fig:Setup}a) transmission. The all-fiber path contains a variable optical attenuator (VOA) used to emulate the loss of the free-space link of $\sim$20 dB. The free-space link, schematically shown in Fig.~\ref{fig:Setup}c, consists of a pair of fiber-to-free-space terminals (Koruza Pro) and a removable rotating phase plate (Lexitek PRP-100) to emulate atmospheric turbulence. In this work, the turbulence emulator is a rotation stage-mounted phase plate comprised of layered polymers with pseudo-Kolmogorov phase designs enclosed between two anti-reflection coated glass plates~\cite{kolmogorov1941local,ebstein2002pseudo}, with a coherence length $r_0$  of 0.5~mm. The terminals at each end of the free-space link consist of a large aperture plano-convex lens and a single-mode fiber, which acts as a filter for higher-order modes. Finally, a second WSS separates the two channels to then be sent to individual coherent receivers which are synchronized with a square pulse generated by an arbitrary waveform generator~(AWG). The received waveforms are processed offline, where the error vector magnitude~(EVM) of the received symbols relative to the ideal symbols is used to monitor the performance of the link.\\
\subsection{Link Characterization}
\begin{figure}[htbp]
\centering\includegraphics[width=0.9\columnwidth]{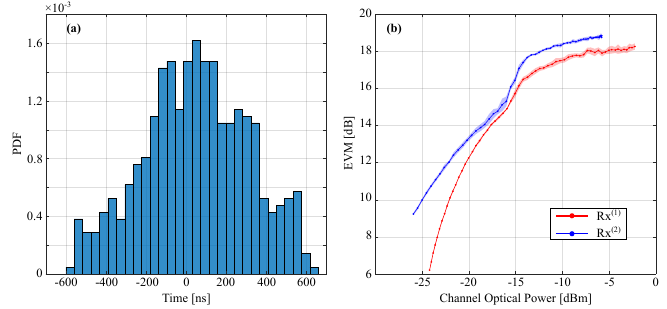}
\caption{Using the all-fiber link in Fig.~\ref{fig:Setup}a path~1, (a) Histogram showing the timing accuracy of the synchronized receivers relying on an arbitrary waveform generator~(AWG). The power of each received waveform from Rx$^{(1)}$ and Rx$^{(2)}$ are cross-correlated over 500 pairs of captures, yielding an average of 48~ns and standard deviation of 263~ns. (b) Error vector magnitude~(EVM) relative to ideal symbols plotted as a function of channel optical power for each coherent receiver. Markers represent the mean and shaded areas represent the standard deviation over 10 measurements.}
\label{fig:PowerCal}
\end{figure}

In order to emulate the hardware implementation of a dual-wavelength receiver, such as in Fig.~\ref{fig:FSOCLink}, that processes each wavelength simultaneously, we ensure the two single-wavelength receivers are synchronized. To do so, the experimental setup is first characterized using the all-fiber path~1 in Fig.~\ref{fig:Setup}a. The synchronization accuracy between receivers is measured by setting the Tx to transmit a pseudo-random binary sequence, where 500 pairs of captured waveforms, each approximately 1.14~$\mu$s in duration, are collected from the two receivers. The power of each waveform is extracted and cross-correlated to retrieve the delay between waveforms at each receiver (Fig.~\ref{fig:PowerCal}a). The result is an average delay of 48~ns and a standard deviation of 263~ns, ensuring that at least one full waveform repetition can be captured during our experiments.\\
To determine the sensitivity of each receiver, using the all-fiber link the received waveforms are processed as the input optical power into the receivers is swept via the VOA. Figure~\ref{fig:PowerCal}b displays the recovered symbol EVM as a function of the input optical power and demonstrates that Rx$^{(1)}$ can receive less than -26~dBm before reaching the required SNR (RSNR) of 7.06~dB while the EVM of Rx$^{(2)}$ falls below the RSNR at an input optical power approaching -24~dBm. Near -15~dBm optical power, a lossy tributary results in a penalty and causes the EVM to decrease rapidly until the other tributaries also experience loss.\\
\subsection{Waveform Combining}\label{section:combining}
A post-processing procedure is used to combine synchronized waveforms from Rx$^{(1)}$ and Rx$^{(2)}$ and achieve MRC, emulating the scenario where the signals are combined in the optical domain~(Fig.~\ref{fig:FSOCLink}).\\
First, the receiver calibration -- transfer function, tributary skews, and quadrature error -- is applied to the respective waveforms $E_{Rx^{(1)}}(t)$ and $E_{Rx^{(2)}}(t)$, which are subsequently polarization-aligned and trimmed to a single repetition of the transmitted waveform. Next, the intermediate frequency is removed and subsample time alignment is performed. The waveforms are then individually processed offline to extract their respective EVMs. We define $\alpha = EVM_{Rx^{(2)}}/EVM_{Rx^{(1)}}$ as a weighting parameter for the combining stage, where the digitally combined waveform is of the form $E_C(t) = (E_{Rx^{(1)}}(t) + \alpha E_{Rx^{(2)}}(t))/(1+\alpha)$. If there is a large EVM difference between the two wavelengths, and therefore a large difference in optical power, $\alpha$ ensures that the influence of the lower power wavelength is representative of the situation where the signals are combined optically. Finally, $E_C(t)$ is processed offline to extract the EVM of the combined signals in the same manner that the individual waveforms $E_{Rx^{(1)}}(t)$ and $E_{Rx^{(2)}}(t)$ are processed.
\section{Results and Discussion}
\begin{figure}[b]
\centering\includegraphics[width=\columnwidth]{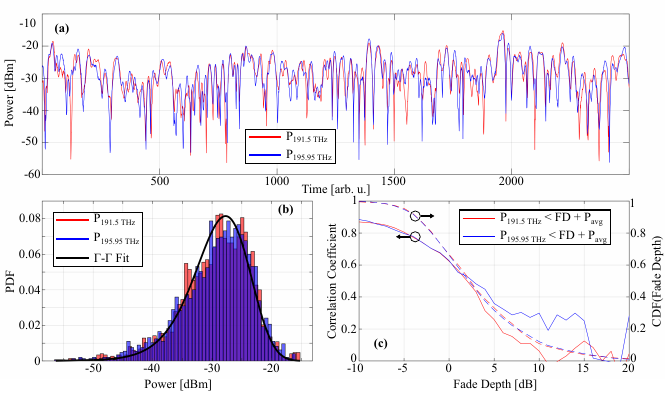}
\caption{(a) Channel powers~($P$) extracted via OSA traces after propagation through the turbulent free-space link, yielding an overall correlation coefficient of 0.9. (b) Received power histograms of each transmitted optical frequency and fit to the Gamma-Gamma model described in Eq.~\ref{eq:gammagamma}. (c) Correlation coefficient (left y-axis, solid lines) between channel powers when the low-frequency wave (red) and high-frequency wave (blue) are at a given fade depth~(FD) relative to the mean power over the entire record shown in (a). Dashed lines correspond to the CDF of the fade depth~(right y-axis) in the correlation coefficient calculations, analogous to the percentage of captures considered in the calculation.}
\label{fig:WavCorr}
\end{figure}
\begin{figure}[h]
\centering\includegraphics[width=0.6\columnwidth]{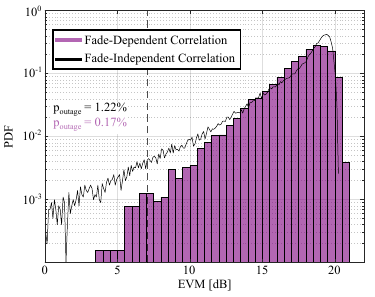}
\caption{Histogram of EVMs, plotted in log scale, after following the waveform combining procedure described in Section~\ref{section:combining}~(purple) and the calculated PDF based on the EVM at individual receivers and the overall power correlation of 0.9~(black line). The dashed black line represents the RSNR of the modulation format.}
\label{fig:CombinedHistogram}
\end{figure}

To extract the power correlation between channels over the turbulent free-space link, optical spectrum analyzer (OSA) traces are gathered as the phase plate is incrementally stepped. The channel power is then calculated from the OSA traces by integrating the power over the signal bandwidth of 56.8~GHz. The optical power of each wavelength over time is shown in Fig.~\ref{fig:WavCorr}a. The received power histograms are shown in Fig.~\ref{fig:WavCorr}b and are fit to the Gamma-Gamma model described in Eq.~\ref{eq:gammagamma}, where $\alpha$ and $\beta$ are used as fitting parameters with values of 4.3 and 2.4, respectively. Channel powers exhibit a correlation coefficient of 0.9 across the entirety of the measurement record. While lower correlation is generally preferred for wavelength diversity, in this case, the degree of decorrelation aligns with the spectral separation of the two wavelengths positioned at the edges of the C-band~(Eq.~\ref{Eq:CoherenceFunction}). However, the correlation decreases substantially in scenarios where one or both channels are subjected to power fading. In Fig.~\ref{fig:WavCorr}c, the correlation coefficient between channel powers is plotted as a function of fade depth (solid lines, left axis). Here, the correlation coefficient is calculated by considering only the traces below a given fade depth relative to the mean power of the entire record. As fade depths exceed a few dB, the correlation coefficient between channel powers decreases significantly, indicating that as one channel fades, the other does not necessarily fade concurrently. As a result, the low-power channel, which also exhibits a low EVM, does not dramatically affect the EVM of the combined signal if MRC is performed. Additionally, Fig.~\ref{fig:WavCorr}c displays the number of captures considered in the correlation coefficient calculations (dashed lines, right axis).\\
\begin{figure}[b]
\centering\includegraphics[width=\columnwidth]{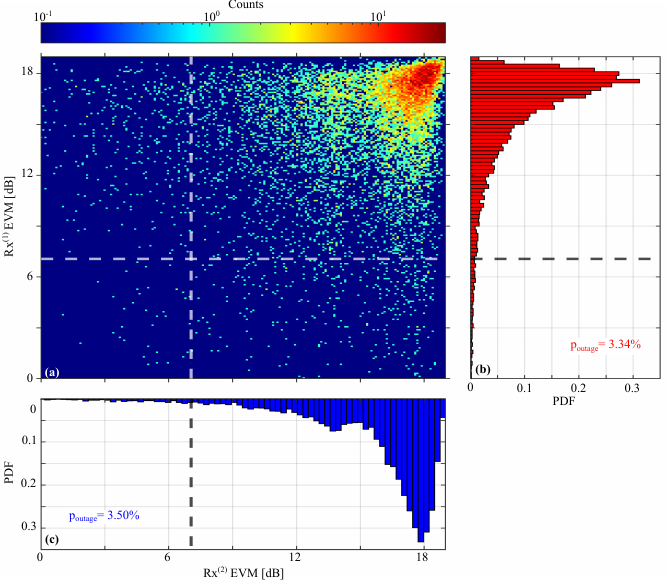}
\caption{(a) Binned scatter plot displaying Rx$^{(1)}$ and Rx$^{(2)}$ EVMs over the turbulent free-space channel, including 12.7k pairs of captures. Histograms showing the EVM distribution of (b) Rx$^{(2)}$ and (c) Rx$^{(1)}$. The individual channels have an outage probability greater than 3\%. In all sub-figures, the dashed lines indicate the RSNR of the modulation format.}
\label{fig:Histograms}
\end{figure}12.7k waveform pairs were captured and processed after propagation over the turbulent link emulator, with the resulting EVM of each channel displayed in the scatter plot of Fig.~\ref{fig:Histograms}a. Due to the high correlation in the optical powers of the channels, the majority of captures cluster near the high-EVM regime around 18~dB, where the received optical power of both wavelengths exceeds the lock point of the receivers. However, there are numerous instances where one channel falls below the RSNR (dashed lines in Fig.~\ref{fig:Histograms}) while the other remains above it -- though significantly fewer instances ($\sim$0.3\%) occur where both channels fall below RSNR. Figures \ref{fig:Histograms}b and \ref{fig:Histograms}c illustrate the EVM probability density function~(PDF) observed at Rx$^{(1)}$ and Rx$^{(2)}$, respectively. In both cases, the distributions peak near an EVM of 18~dB, exhibit a maximum EVM of 19~dB, and feature a tail extending down to 0~dB. The resulting outage probabilities are 3.5\% for Rx$^{(1)}$ and 3.34\% for Rx$^{(2)}$.\\
A histogram of the resulting EVM after MRC of waveforms is shown in Fig.~\ref{fig:CombinedHistogram}, following the steps in Section~\ref{section:combining}. In this case, the distribution peaks near 19~dB, reaches a maximum EVM of 20.8~dB, and features a tail extending down to 3.7~dB. Notably, the outage probability drastically decreased to 0.17\%, representing an improvement of nearly 20 times compared to single-wavelength transmission over the link. 
This experimental result is compared with the expected PDF after MRC, shown as a black line in Fig.~\ref{fig:CombinedHistogram}. This expected PDF is calculated by generating 100k correlated random variables from the single-wavelength distributions shown in Figs.~\ref{fig:Histograms}b and c, with the assumption that there is no fade-dependent correlation. The overall power correlation between waves of 0.9 is therefore used for the calculation. Under these conditions, the expected outage probability remains elevated at 1.22\%, more than 7 times higher than in our experimental results where there is a fade-dependent correlation.\\
\section{Conclusion}
In summary, we have experimentally demonstrated wavelength diversity over an emulated turbulent free-space channel using a dual-wavelength coherent transceiver operating at 56.8~GBaud and 200~Gbps. This approach has significantly improved the outage probability, by nearly a factor of 20, compared to traditional single-wavelength transmission. These results highlight the potential of wavelength diversity to enhance the reliability of free-space optical communication links in environments affected by strong atmospheric turbulence. Additionally, our experiments revealed a regime of elevated decoherence that occurs when both wavelengths experience a deep fade, despite the overall correlation between the wavelengths being high. The recorded outage probability of 0.17\% in our measurements is 86\% lower than the expected outage probability of 1.22\%, which is calculated with an MRC approach with a fade-independent correlation coefficient of 0.9. By leveraging the inherent robustness of redundant information spread across multiple wavelengths, this technique offers a practical way to improve communication reliability. This advancement paves the way for widespread adoption of FSOC technology in applications such as satellite communications, urban connectivity, and even disaster recovery scenarios.

\begin{backmatter}
\bmsection{Acknowledgments}
The authors would like to extend their sincere gratitude to the Ciena research and development team for providing access to WaveLogic~Ai modems. Furthermore, the authors acknowledge Shahab~Oveis~Gharan, Charles~Laperle, Jiaying~Lu, Andrzej~Borowiec, Maurice~O’Sullivan, and Michael~Reimer for technical assistance and insightful discussions.

\bmsection{Disclosures}
US Patent App. 18/541,189

\bmsection{Data availability} Data underlying the results presented in this paper are not publicly available at this time but may be obtained from the authors upon reasonable request.

\end{backmatter}

\bibliography{References}

\end{document}